
\font\twelverm=cmr10 scaled 1200    \font\twelvei=cmmi10 scaled 1200
\font\twelvesy=cmsy10 scaled 1200   \font\twelveex=cmex10 scaled 1200
\font\twelvebf=cmbx10 scaled 1200   \font\twelvesl=cmsl10 scaled 1200
\font\twelvett=cmtt10 scaled 1200   \font\twelveit=cmti10 scaled 1200
\font\twelvesc=cmcsc10 scaled 1200  
\skewchar\twelvei='177   \skewchar\twelvesy='60
     
     
\def\twelvepoint{\normalbaselineskip=12.4pt plus 0.1pt minus 0.1pt
  \abovedisplayskip 12.4pt plus 3pt minus 9pt
  \belowdisplayskip 12.4pt plus 3pt minus 9pt
  \abovedisplayshortskip 0pt plus 3pt
  \belowdisplayshortskip 7.2pt plus 3pt minus 4pt
  \smallskipamount=3.6pt plus1.2pt minus1.2pt
  \medskipamount=7.2pt plus2.4pt minus2.4pt
  \bigskipamount=14.4pt plus4.8pt minus4.8pt
  \def\rm{\fam0\twelverm}          \def\it{\fam\itfam\twelveit}%
  \def\sl{\fam\slfam\twelvesl}     \def\bf{\fam\bffam\twelvebf}%
  \def\mit{\fam 1}                 \def\cal{\fam 2}%
  \def\sc{\twelvesc}               \def\tt{\twelvett}
  \def\sf{\twelvesf}
  \textfont0=\twelverm   \scriptfont0=\tenrm   \scriptscriptfont0=\sevenrm
  \textfont1=\twelvei    \scriptfont1=\teni    \scriptscriptfont1=\seveni
  \textfont2=\twelvesy   \scriptfont2=\tensy   \scriptscriptfont2=\sevensy
  \textfont3=\twelveex   \scriptfont3=\twelveex  \scriptscriptfont3=\twelveex
  \textfont\itfam=\twelveit
  \textfont\slfam=\twelvesl
  \textfont\bffam=\twelvebf \scriptfont\bffam=\tenbf
  \scriptscriptfont\bffam=\sevenbf
  \normalbaselines\rm}
     

     
\def\beginlinemode{\endmode
  \begingroup\parskip=0pt \obeylines\def\\{\par}\def\endmode{\par\endgroup}}
\def\beginparmode{\endmode
  \begingroup \def\endmode{\par\endgroup}}
\let\endmode=\par
{\obeylines\gdef\
{}}
\def\singlespace{\baselineskip=\normalbaselineskip}

\def\oneandahalfspace{\baselineskip=\normalbaselineskip
  \multiply\baselineskip by 3 \divide\baselineskip by 2}
\def\doublespace{\baselineskip=\normalbaselineskip \multiply\baselineskip by 2}

\newcount\firstpageno
\firstpageno=2
\footline={\ifnum\pageno<\firstpageno{\hfil}\else{\hfil\twelverm\folio\hfil}\fi}
\def\toppageno{\global\footline={\hfil}\global\headline
  ={\ifnum\pageno<\firstpageno{\hfil}\else{\hfil\twelverm\folio\hfil}\fi}}
\let\rawfootnote=\footnote              
\def\footnote#1#2{{\rm\singlespace\parindent=0pt\parskip=0pt
  \rawfootnote{#1}{#2\hfill\vrule height 0pt depth 6pt width 0pt}}}
\def\raggedcenter{\leftskip=4em plus 12em \rightskip=\leftskip
  \parindent=0pt \parfillskip=0pt \spaceskip=.3333em \xspaceskip=.5em
  \pretolerance=9999 \tolerance=9999
  \hyphenpenalty=9999 \exhyphenpenalty=9999 }
\def\dateline{\rightline{\ifcase\month\or
  January\or February\or March\or April\or May\or June\or
  July\or August\or September\or October\or November\or December\fi
  \space\number\year}}
\def\received{\vskip 3pt plus 0.2fill
 \centerline{\sl (Received\space\ifcase\month\or
  January\or February\or March\or April\or May\or June\or
  July\or August\or September\or October\or November\or December\fi
  \qquad, \number\year)}}
     
     
\hsize=6.5truein
\vsize=8.5truein
\parskip=\medskipamount
\def\\{\cr}
\twelvepoint            
\doublespace            
\overfullrule=0pt       

\def\title                      
  {\null\vskip 3pt plus 0.2fill
   \beginlinemode \doublespace \raggedcenter \bf}
     
\def\author                     
  {\vskip 3pt plus 0.2fill \beginlinemode
   \singlespace \raggedcenter\sc}
     
\def\affil                      
  {\vskip 3pt plus 0.1fill \beginlinemode
   \oneandahalfspace \raggedcenter \sl}
     
\def\abstract                   
  {\vskip 3pt plus 0.3fill \beginparmode
   \singlespace ABSTRACT: }
     
\def\endtopmatter               
  {\endpage                     
   \body}
     
\def\body                       
  {\beginparmode}               
     
\def\head#1{                    
  \goodbreak\vskip 0.5truein    
  {\immediate\write16{#1}
   \raggedcenter \uppercase{#1}\par}
   \nobreak\vskip 0.25truein\nobreak}

\def\beginitems{
\par\medskip\bgroup\def\i##1 {\item{##1}}\def\ii##1 {\itemitem{##1}}
\leftskip=36pt\parskip=0pt}
\def\enditems{\par\egroup}
     
\def\beneathrel#1\under#2{\mathrel{\mathop{#2}\limits_{#1}}}
     
\def\refto#1{$^{#1}$}           
     
\def\references                 
  {\head{References}            
   \beginparmode
   \frenchspacing \parindent=0pt \leftskip=1truecm
   \parskip=8pt plus 3pt \everypar{\hangindent=\parindent}}

\gdef\refis#1{\item{#1.\ }}                     
     
\gdef\journal#1, #2, #3, 1#4#5#6{               
    {\sl #1~}{\bf #2}, #3 (1#4#5#6)}            

\gdef\refa#1, #2, #3, #4, 1#5#6#7.{\noindent#1, #2 {\bf #3}, #4 (1#5#6#7).\rm} 

\gdef\refb#1, #2, #3, #4, 1#5#6#7.{\noindent#1 (1#5#6#7), #2 {\bf #3}, #4.\rm} 

\def\pr{\journal Phys.Rev., }

\def\prl{\journal Phys.Rev.Lett., }

\def\cmp{\journal Comm.Math.Phys., }
     
\def\np{\journal Nucl.Phys., }

\def\annp{\journal Ann.Phys.(N.Y.), }

\def\endreferences{\body}

\def\endpage                    
  {\vfill\eject}
     
\def\endpaper                   
  {\endmode\vfill\supereject}

\def\ref#1{Ref.~#1}                     
\def\Ref#1{Ref.~#1}                     
\def\[#1]{[\cite{#1}]}
\def\cite#1{{#1}}
\def\(#1){(\call{#1})}
\def\call#1{{#1}}
\def\taghead#1{}
\def\frac#1#2{{#1 \over #2}}
\def\half{{\frac 12}}

\def\12{{1\over2}}

\catcode`@=11
\newcount\r@fcount \r@fcount=0
\newcount\r@fcurr
\immediate\newwrite\reffile
\newif\ifr@ffile\r@ffilefalse
\def\w@rnwrite#1{\ifr@ffile\immediate\write\reffile{#1}\fi\message{#1}}

\def\writer@f#1>>{}
\def\referencefile{
  \r@ffiletrue\immediate\openout\reffile=\jobname.ref%
  \def\writer@f##1>>{\ifr@ffile\immediate\write\reffile%
    {\noexpand\refis{##1} = \csname r@fnum##1\endcsname = %
     \expandafter\expandafter\expandafter\strip@t\expandafter%
     \meaning\csname r@ftext\csname r@fnum##1\endcsname\endcsname}\fi}%
  \def\strip@t##1>>{}}

\def\citeall#1{\xdef#1##1{#1{\noexpand\cite{##1}}}}
\def\cite#1{\each@rg\citer@nge{#1}}	

\def\each@rg#1#2{{\let\thecsname=#1\expandafter\first@rg#2,\end,}}
\def\first@rg#1,{\thecsname{#1}\apply@rg}	
\def\apply@rg#1,{\ifx\end#1\let\next=\relax
\else,\thecsname{#1}\let\next=\apply@rg\fi\next}

\def\citer@nge#1{\citedor@nge#1-\end-}	
\def\citer@ngeat#1\end-{#1}
\def\citedor@nge#1-#2-{\ifx\end#2\r@featspace#1 
  \else\citel@@p{#1}{#2}\citer@ngeat\fi}	
\def\citel@@p#1#2{\ifnum#1>#2{\errmessage{Reference range #1-#2\space is bad.}%
    \errhelp{If you cite a series of references by the notation M-N, then M and
    N must be integers, and N must be greater than or equal to M.}}\else%
 {\count0=#1\count1=#2\advance\count1 by1\relax\expandafter\r@fcite\the\count0,
  \loop\advance\count0 by1\relax
    \ifnum\count0<\count1,\expandafter\r@fcite\the\count0,%
  \repeat}\fi}

\def\r@featspace#1#2 {\r@fcite#1#2,}	
\def\r@fcite#1,{\ifuncit@d{#1}
    \newr@f{#1}%
    \expandafter\gdef\csname r@ftext\number\r@fcount\endcsname%
                     {\message{Reference #1 to be supplied.}%
                      \writer@f#1>>#1 to be supplied.\par}%
 \fi%
 \csname r@fnum#1\endcsname}
\def\ifuncit@d#1{\expandafter\ifx\csname r@fnum#1\endcsname\relax}%
\def\newr@f#1{\global\advance\r@fcount by1%
    \expandafter\xdef\csname r@fnum#1\endcsname{\number\r@fcount}}

\let\r@fis=\refis			
\def\refis#1#2#3\par{\ifuncit@d{#1}
   \newr@f{#1}%
   \w@rnwrite{Reference #1=\number\r@fcount\space is not cited up to now.}\fi%
  \expandafter\gdef\csname r@ftext\csname r@fnum#1\endcsname\endcsname%
  {\writer@f#1>>#2#3\par}}

\def\ignoreuncited{
   \def\refis##1##2##3\par{\ifuncit@d{##1}%
    \else\expandafter\gdef\csname r@ftext\csname r@fnum##1\endcsname\endcsname%
     {\writer@f##1>>##2##3\par}\fi}}

\def\r@ferr{\endreferences\errmessage{I was expecting to see
\noexpand\endreferences before now;  I have inserted it here.}}
\let\r@ferences=\references
\def\references{\r@ferences\def\endmode{\r@ferr\par\endgroup}}

\let\endr@ferences=\endreferences
\def\endreferences{\r@fcurr=0
  {\loop\ifnum\r@fcurr<\r@fcount
    \advance\r@fcurr by 1\relax\expandafter\r@fis\expandafter{\number\r@fcurr}%
    \csname r@ftext\number\r@fcurr\endcsname%
  \repeat}\gdef\r@ferr{}\endr@ferences}


\let\r@fend=\endpaper\gdef\endpaper{\ifr@ffile
\immediate\write16{Cross References written on []\jobname.REF.}\fi\r@fend}

\catcode`@=12

\citeall\refto		
\citeall\ref		%
\citeall\Ref		%

\def\a{{\alpha}}

\def\s{\sigma}
\def\half{{1 \over 2}}
\def\ra{{\rangle}}
\def\la{{\langle}}

\def\ih{{i \over \hbar}}

\def\q{{\bar q}}

\def\x{{\bar x}}

\def\D{{\cal D}}

\def\ria{{\rightarrow}}

\centerline{\bf Effective Theories of Coupled}
\centerline{\bf Classical and Quantum Variables}

\vskip 0.3in

\author{J. J. Halliwell} 
\affil
Theory Group
Blackett Laboratory
Imperial College 
London SW7 2BZ
UK
\vskip 0.2in
\centerline{\rm Preprint Imperial/TP/97-98 /70, August 1998}
\vskip 0.5in
{\rm To appear in proceedings of the conference,}
{\it Open Systems and Measurement in Relativistic Quantum Theory},
{\rm Naples, April 3-4, 1998, edited by H.P.Breuer and F.Petruccione}
\vskip 0.2in

\abstract{We address the issue of coupling variables which are
essentially classical to variables that are quantum. Two approaches
are discussed. In the first (based on collaborative work with
L.Di\'osi), continuous quantum measurement theory is used to
construct a phenomenological description of the interaction of a
quasiclassical variable $X$ with a quantum variable $x$, where the
quasiclassical nature of $X$ is assumed to have come about as a
result of decoherence. The state of the quantum subsystem evolves
according to the stochastic non-linear Schr\"odinger equation of a
continuously measured system, and the classical system couples to a
stochastic c-number $\x (t)$ representing the imprecisely measured
value of $x$. The theory gives intuitively sensible results even
when the quantum system starts out in a superposition of
well-separated localized states. The second approach involves a
derivation of an effective theory from the underlying quantum theory
of the combined quasiclassical--quantum system, and uses the
decoherent histories approach to quantum theory.
}

\endtopmatter
\endpage

\head{\bf 1. Introduction}

What happens when a classical system interacts with a quantum system
in a non-trivial superposition state? Quantum field theory in curved
spacetime is an example of a number of situations where one would
like to know the answer to this question. There, the effect of the
quantized matter field on the classical gravitational field is
often assessed using the semiclassical Einstein equations
[\cite{Ros,Mol}]:
$$
G_{\mu \nu} = 8 \pi G \la T_{\mu \nu} \ra 
\eqno(1.1)
$$
The left hand side is the Einstein tensor of the classical
metric field $g_{\mu \nu} $ and the right hand side is the
expectation value of the energy momentum tensor of a quantum
field. 

Although we do not yet have the complete, workable theory of quantum
gravity required to derive an equation like (1.1), on general
grounds it is clear that it is unlikely to be valid unless the
fluctuations in $T_{\mu\nu} $ are small [\cite{For,HaH,KuF}].
Indeed, (1.1) fails to give intuitively sensible results when the
matter field is in a superposition of localized states 
[\cite{PaG,Kib1}].  In particular, when the quantum state of the
matter field consists of a superposition of two well-separated
localized states, Eq.(1.1) suggests that the gravitational field
couples to the average energy density of the two states, whilst
physical intuition suggests that the gravitational field feels the
energy of one or other of the localized matter states, with some
probability.  It is by no means obvious,  however, that we have to
resort to quantum gravity to accommodate such non-trivial matter states.
This leads one to ask whether there exists a semiclassical theory
with a much wider range of validity than (1.1), which gives
intuitively reasonable results for non-trivial superposition states
for the matter field.

The aim of this contribution is to describe two related approaches to
coupling classical and quantum variables which go far beyond the
naive mean field equations, and produce intuitively sensible results
in the key case of superposition states.  The full problem of the
semiclassical Einstein equations (1.1) will not be addressed.
Rather, we will concentrate on a simple model in which the scheme is
easily presented and perhaps verified.  Of course, many previous
authors have tackled this problem [\cite{And,Ale,Dio2,Zou}].
What is perhaps new in the present approach compared to previous
ones is the explicit incorporation of the notion of decoherence
to ensure that the ``classical'' system really is classical.
(See, however, Ref.[\cite{And}], for some earlier comments
along these lines.)

Our considerations will be based entirely on the following simple
model, consisting of a classical particle with position $X$ in a
potential $V(X)$ coupled to a  harmonic oscillator with position $x$
which will later be quantized.  The action is
$$
S= \int dt \ \left( \half M \dot X^2 - V(X) + \half m \dot x^2 - \half m
\omega^2 x^2 - \lambda X x \right) .
\eqno(1.2) 
$$ 
Hence the classical equations of motion are 
$$ 
\eqalignno{ 
M \ddot X + V'(X) + \lambda x &= 0 ,
&(1.3)\cr 
m \ddot x + m \omega^2 x + \lambda X &= 0 . 
&(1.4) \cr }
$$
The naive mean field approach involves replacing (1.3) with the
equation
$$
M \ddot X + V'(X) + \lambda \la \psi | \hat x | \psi \ra = 0 ,
\eqno(1.5)
$$
and replacing (1.4) with the Schr\"odinger equation 
$$
{d \over dt} | \psi \ra = - \ih \left( 
\hat H_0 + \lambda X \hat x \right) | \psi \ra 
\eqno(1.6)
$$
for the quantum particle. $\hat H_0$ is the Hamiltonian of the
quantum particle (in this case a harmonic oscillator) and $-X(t)$ is
regarded as an external classical force. As stated above, the scheme
(1.5), (1.6) is unlikely to have a very wide range of validity.

Generally, for a quantum system with wave function $\psi (x) $,
there will be a non-zero probability for $x$ to take any one of a 
range of values, and the expectation value $ \la \hat x \ra $  (as
in Eq.(1.5)) will not be representative of the distribution of $x$
(unless the distribution just happens to be peaked about its
expectation value). One would therefore expect the classical system
to be stochastically influenced by the quantum system and follow
one of an ensemble trajectories.
To be precise, we expect an improved version of
(1.5) to be of the form
$$
M \ddot X + V'(X) + \lambda \bar x (t) = 0 ,
\eqno(1.7)
$$
where $\bar x (t) $ is now a classical {\it stochastic} variable,
whose probability distribition is determined by the dynamics and
quantum state of the quantum particle. 

The purpose of this paper is to describe two different but related
approaches to coupling classical and quantum variables, both of
which lead to an equation of the form (1.7) and both of which yield
an explicit probability distribution for $\x (t) $. The first
approach (which was developed in collaboration with Lajos Di\'osi) 
is a phenomenological scheme based on continuous quantum measurement
theory. The second is a more fundamentally based scheme, derived
using the decoherent histories approach to quantum theory. This work
is based on two published papers [\cite{DiH,Hal3}].

Taking the second of these schemes first,
the question of coupling classical variables to quantum variables
is intimately connected to the question of how certain
variables become classical in the first place. We
adopt the point of view that there are no  {\it fundamentally}
classical systems in the world, only quantum systems that are
effectively classical under certain conditions. The most
comprehensive approach to obtaining generalizations of the
semiclassical scheme (1.5), (1.6), therefore consists of starting from 
the underlying quantum theory of the whole composite system, and
then {\it deriving} the effective form of that theory under the
conditions in which one of the subsystems is effectively classical. 
The most important condition that needs to be satisfied for a
subsystem to be effectively classical is {\it decoherence} --
interference between histories of certain types of variables (in
this case position) must be destroyed (see, for example,
Ref.[\cite{GeH1,GeH2}]). Decoherence is typically brought about
by some kind of coarse-graining procedure, of which perhaps the most
commonly used procedure is to couple to a large environment
(typically a heat bath) and then trace it out. The resulting
decoherent variables are often referred to as quasiclassical, a
nomenclature we shall adopt. Quasiclassical variables follow
classical trajectories but modified by fluctuations induced by the
environment that decoherered them. For sufficiently massive
particles, these fluctuations have negligible effect. 

A derivation of an effective theory of coupled quasiclassical and
quantum variables therefore involves a three-component quantum
system consisting of a (``to be quasiclassical'') particle with
position $X$, coupled to an environment which is traced out to
render $X$ quasiclassical, and also coupled to the position $x$ of
another (``quantum'') particle (not necessarily coupled to the
environment). In Section 3, we will show, in the context
of a particular model, how such an effective theory may be derived
using the decoherent histories approach to quantum theory.

Emergent classicality is, however, a widespread and generic
phenomenon. It has been demonstrated in a wide variety of different
models using a variety of different approaches to decoherence. This
suggests that it ought to be possible to directly write down a
phenomenological model describing the coupling of the quasiclassical
variable $X$ to the quantum variable $x$,  but without having to
appeal to the full details of a specific decoherence calculation. 
Differently put, it is of interest to determine the minimal
elaboration required of Eqs. (1.5), (1.6), to obtain a viable scheme of
coupled classical--quantum variables. Such a scheme
would also have the advantage that it may be valid when the
underlying quantum theory is not particularly manageable or even not
known (as may be the case for gravity). 

For these reasons, in Section 2, a more phenomenological
approach to classical quantum couplings is presented.
This approach is based on
the observation that there already exists a partial description of
classical--quantum couplings in the form of continuous quantum
measurement theory. This existing structure, together with a
heurisitic appreciation of decoherence leads to the desired
phenomenological scheme.  The idea is to think of the 
quasiclassical particle as in some sense ``measuring'' the quantum
particle's position and responding to the measured c-number result
$\x$. (A precursor to this idea may be found in Ref.[\cite{Dio3}]). 
In this approach, the decoherence of the quasiclassical particle is
not modeled explicitly, but an appeal is made to general known
features of the decoherence process where necessary. In particular,
the assumed decoherence ensures that the quasiclassical particle
remains quasiclassical  (although may be stochastically influenced)
even when it interacts with the quantum particle in a non-trivial
superposition.

The two models are summarized in Section 4.

\head{\bf 2. Classical--Quantum Couplings via
Continuous Quantum Measurement Theory}

As stated in the Introduction, the first approach to coupling
classical and quantum variables is a phenomenological scheme using
continuous quantum measurement theory. The basic idea is to think
of the classical variable as in some sense ``measuring'' the quantum
particle and responding to the measured c-number result.

Consider, therefore, the consequences of standard quantum
measurement theory for the evolution of the coupled quasiclassical
and quantum systems over a small interval of time $\delta t $. The
state $| \psi \ra $ of the quantum system will evolve, as a result
of the measurement, into the (unnormalized) state
$$
| \Psi_{\x} \ra = \hat P_{\x}  e^{- i \hat H \delta t} | \psi \ra
\eqno(2.1)
$$
where $\hat H = \hat H_0 + \lambda X \hat x $  and $\hat P_{\x}$ is
a projection operator which asks whether the position of the quantum
particle is $\x$, to within some precision. (If the classical system
couples to some operator of the quantum system other than position,
{\it e.g.}, momentum, then the projection operator in (2.1) is changed
accordingly, {\it e.g.}, to a momentum projector). The probability
that the measurement yields the result $\x$ is given by $ \la
\Psi_{\x} | \Psi_{\x} \ra $. It is then natural to suppose that the
classical particle, in responding to the measured result, will
evolve during this small time interval according to the equation of
motion
$$
M \ddot X + V'(X) + \lambda \x  = 0,
\eqno(2.2)
$$
with probability $ \la \Psi_{\x} | \Psi_{\x} \ra $.

Now we would like to 
repeat the process for an arbitrary number of time steps and
then take the continuum limit. If $\hat P_{\x}$ is
an exact projection operator, {\it i.e.}, one for which 
$\hat P^2_\x = \hat P_\x$, the continuum limit is trivial and 
of no interest (this is the watchdog effect). However, standard
quantum measurement theory has been generalized to a well-defined
and non-trivial
process that acts continuously in time by replacing $\hat P_{\x}$ with a
positive operator-valued measure (POVM) [\cite{Dio3,BLP,CaM,Dio4}]. 
The simplest example, which we use here, is a Gaussian,
$$
\hat P_{\x} = {1 \over ( 2 \pi \Delta^2 )^{\half} }
\ \exp \left( - { ( \hat x - \x )^2
\over 2 \Delta^2 } \right) 
\eqno(2.3)
$$
and the continuum limit involves taking $\Delta \ria \infty $ as
$ \delta t \ria 0 $ in such a way that $\Delta^2 \delta t $ is held
constant. 
The evolution of the wave function
of the quantum system is then conveniently expressed in terms of a
path-integral expression for the unnormalized wave function:
$$
\eqalignno{
\Psi_{[\x (t)]} (x',t') = \int \D x
& \exp \left( \ih \int_0^{t'} dt \left( \half m \dot x^2 - \half m \omega^2
x^2 - \lambda x X \right) \right)
\cr
\ \times \exp & \left( - 
\int_0^{t'} dt { ( x - \x )^2 \over 4 \s^2 } \right)
\ \Psi (x_0, 0 ).
&(2.4) \cr}
$$
Here, the integral is over paths $ x(t) $ satisfying $x(0) = x_0 $
and $x(t') = x' $.
The classical particle at each moment of time evolves according to
Eq.(2.2), where the functional probability distribution of the entire
measured path $\x(t)$ takes the form:
$$
p [\x (t) ] = \la \Psi_{[ \x (t) ]} | \Psi_{[ \x (t) ]} \ra.
\eqno(2.5)
$$
(The parameter $\s$ in Eq.(2.5), representing the width of
the effective ``measurement'' of the particle by the classical
system, will be discussed below).

The scheme is therefore as follows. We solve the equations (2.2) and
(2.4) where $\x (t) $ is regarded as a stochastic variable whose
probability distribution is given by (2.5). The final result is
an ensemble of $\x$-dependent classical and quantum
trajectories respectively for the  two particles, with an
interdependent probability distribution. 

It turns out that this system (2.2), (2.4), (2.5) can be rewritten in
such a way that brings it closer to the form of the naive mean field
equations (1.5), (1.6). The basic issue is that Eq.(2.5) gives the
probability for an entire history of measured alternatives, $\x
(t)$. Yet the naive mean field equations (1.5), (1.6) are evolution
equations defined at each moment of time.  Fortunately, the system (2.2),
(2.4), (2.5) may be rewritten as follows. Consider the basic process
(2.1) with the Gaussian projector (2.3), but in addition let the 
state vector be normalized at each time step.  Then denoting the
normalized state at each time by $ | \psi \ra $, and taking the
continuum limit in the manner indicated above, it is readily shown
[\cite{Dio4}] that $ | \psi \ra $ obeys a stochastic non-linear
equation describing a system undergoing continuous measurement:
$$
{d \over dt} | \psi \ra = \left( - \ih ( \hat H_0 + \lambda X \hat x ) 
- {1 \over  4 \s^2} 
( \hat x - \la \hat x \ra )^2 \right) | \psi \ra 
+ {1 \over 2 \s } 
\left( \hat x - \la \hat x \ra \right) | \psi \ra \eta (t).
\eqno(2.6)
$$
Here, $\eta (t)$ is the standard Gaussian white noise, 
with linear and quadratic means,
$$ 
M \left[ \eta (t) \right] = 0, \quad \quad
M \left[ \eta (t) \eta (t') \right]  = \delta (t-t').
\eqno(2.7)
$$
where $ M ( \cdots ) $ denotes stochastic averaging.
The noise terms are to be interpreted in the sense of Ito.
The measured value $\x$ is then related to $\eta $ by
$$
\x = \la \psi | \hat x | \psi \ra + \s  \eta (t). 
\eqno(2.8)
$$
Hence the final form of Eq.(2.2) (replacing Eq.(1.5)), is
$$
M \ddot X + V'(X) + \lambda \la \psi | 
\hat x | \psi \ra + \lambda \s \eta (t)  = 0
\eqno(2.9)
$$
and (1.6) is replaced by the stochastic non-linear equation (2.6).

Turn now to the question of the value of the parameter $\s$.  As
discussed above, the quasiclassical particle suffers fluctuations as
a result of interacting with the environment that decohered it. 
This must still be true even when it is not coupled to the quantum
particle. We can therefore fix $\s$ by demanding that in Eq.(2.9),
the term $\lambda \s \eta (t) $, in the limit $\lambda \ria 0 $,
describe the environmentally induced fluctuations suffered by the
classical particle. This forces us to choose $\s$ to be proportional
to $\lambda^{-1}$. Further information on the form of $\s$ requires
more specific details about the environment. In the particular but
frequently studied case of a thermal environment, the random force
should be $\sqrt{2 M \gamma k_B T}\eta(t)$, in order to coincide
with the standard Langevin equation of classical Brownian motion.
From this we deduce that $\s^2 = 2 M \gamma k_B T / \lambda^2 $.
The result is not hard to
understand. Because of the environmentally induced fluctuations it
suffers, the quasiclassical particle is necessarily limited in the
precision with which it can ``measure'' the quantum particle, hence
the width $\s$ of the ``measurement'' is related to the fluctuations
of the quasiclassical particle.

The formal solution to (2.6) describes a family of pure states,  
$ |\psi \ra = | \psi_{[\eta (t) ] } \ra $,  one for each choice of
function, $\eta (t)$. Correspondingly in Eq.(2.9), with $| \psi \ra =
| \psi_{[\eta (t) ] } \ra $ inserted in the pure state expectation
value, there is one evolution equation for each $\eta (t)$.   For
fixed initial data, $ | \psi_0 \ra $, $X_0$, $\dot X_0$, Eqs.(2.6)
and (2.9) therefore describe an ensemble of quantum and classical
trajectories $ ( | \psi_{[\eta(t)]} \ra, X_{[\eta(t)]}(t))$,  with
members labeled by $\eta (t) $. The probability for each member of
the ensemble is that implied by the probability distribution of
$\eta (t) $ (implicit in Eq.(2.7)).

There are two differences between the system, (2.6)--(2.9) and the
naive mean field equations, (1.5), (1.6).  One is the noise term $\eta$.
In Eq.(2.9) (as compared to Eq.(1.5)) the noise clearly describes an
additional (completely uncorrelated) random force. This sort of
modification to the semiclassical Einstein equations has been
considered previously [\cite{KuF,HuM}].

More important is the novelty that the state $|\psi \ra $ evolves
according to the stochastic non-linear equation (2.6), and hence its
evolution is very different to that under the usual Schr\"odinger
equation, (1.6). 
In particular, it may be shown that all solutions to (2.6)
undergo {\it localization} [\cite{Dio5,Per,GiP,HZ1}] on a
time scale which might be extremely short compared to the
oscillator's freqency $\omega$. That is, every initial state rapidly
evolves to a generalized coherent state centred around values $ \la
\hat x \ra $,  $ \la \hat p \ra $ undergoing classical Brownian
motion. (The results cited above are readily extended to the case
here in which the Hamiltonian contains a linear coupling to an
external force $-X(t)$). Which particular solution the state becomes
centred around depends  statistically on the initial state of the
system. For an initial state consisting of a superposition of
well-separated coherent states,
$$
| \psi \ra = \a_1 | x_1 p_1 \ra + \a_2 | x_2 p_2 \ra
\eqno(2.10)
$$
the state after localization time will, with probability $| \a_1 |^2 $,
be as if the initial state were just $ | x_1 p_1 \ra $, and
with probability $| \a_2 |^2 $, will be as if the initial state were
just $ |x_2 p_2 \ra $ [\cite{HZ1}].
The localization time $\sim 1 / \s^2 (x_1-x_2)^2 $ 
becomes, with our previous choice $\s^2 \sim  M \gamma k_B T / \lambda^2$, 
very short indeed if the classical particle has a large 
mass $M$.

Hence in the new semiclassical equations (2.6)--(2.9), effectively
what happens is that  we solve separately for the two initial states
$| x_1 p_1 \ra $ and $| x_2 p_2 \ra $, and the  classical  particle
then follows the first solution with probability $ | \a_1 |^2 $ and
the second with probability $ | \a_2 |^2 $.
In simple terms, therefore, an almost classical system interacting
through position with a  quantum system in a superposition state
(2.10), ``sees'' one or other of the superposition states, with some
probability, and not the mean position of the entire state. This is
the key case for which the naive mean field equations fail to give
intuitively sensible results [\cite{PaG,Dio}],
and this is the main result of the model.

It is interesting to note that non-linear Schr\"odinger equations
have been considered before in the context of the semiclassical
Einstein equations [\cite{Kib1,Kib2}], because the combined 
system consiting of (1.1) together with the Schr\"odinger equation for
the quantum state is non-linear. The motivation here is rather
different. The equation (2.6) used here arises because it gives a
phenomenological description of continuous measurement. 

Note that our classical stochastic equations (2.9) do not involve
dissipation, as one might expect. Dissipation will arise in Eq.(2.9)
if the model of the measurement process, Eqs.(2.1), (2.3), is extended
to include feedback forces (see, for example, Ref.[\cite{CaM}]).
This would modify our scheme but does not alter it in a fundamental
way. Here, for brevity, we have worked in the commonly used and
instructive approximation of negligible dissipation.

\head {\bf 3. Derivation of an Effective Theory from Decoherent Histories}

In this Section we discuss a more specific but more fundamental
theory of classical--quantum couplings, which is {\it derived}
from the underlying quantum theory of the whole composite
system. As argued in the Introduction, classicality of
a particle arises as a result of decoherence due to the
interaction with an environment. We therefore consider
a three-component composite system consisting of a particle
with coordinate $X$ (eventually to be the quasi-classical system),
coupled to an environment consisting of an infinite number
of harmonic oscillators with coordinates $q_n$ in a thermal state.
The classical particle is also coupled to a small quantum particle
with coordinate $x$. The total action is
$$
\eqalignno{
S & = \int dt \left( \half M \dot X^2 - V(X) \right)
\cr
& + \int dt \sum_n \left( \half m_n \dot q_n^2 - \half m_n
\omega_n^2 q_n^2 + c_n q_n X \right)
\cr
& + \int dt \left( \half m \dot x^2 - \half m \omega^2 x^2 - \lambda
X x \right)
&(3.1) \cr }
$$

We will analyse this system using the decoherent histories approach
to quantum theory. This approach is reviewed in detail elsewhere, so
here, we summarize only the essential parts of it that will be
needed for this calculation
[\cite{GeH1,GeH2,Gri,Hal1,Hal2,Har1,Har2,Omn}]. 

It is not difficult to see why the decoherent histories
approach is useful in this context [\cite{Har3}].
We would like to derive an effective evolution
equation for the variable $X$, which we expect to be approximately
classical motion, plus a stochastic influence from the quantum
system it couples to. We can see whether a particle follows such a
trajectory by computing the probability for a history of positions
distributed in time, {\it i.e.}, an object of the form
$
p( X_1, t_1, X_2, t_2, X_3, t_3, \cdots )
$
This is the probability that the particle is at the approximate
position $X_1$ at $t_1$, at $X_2$ at $t_2$ and so on. 
Because of quantum interference, probabilities cannot immediately be
assigned to histories. We therefore need a mechanism to produce
decoherence of the particle, hence the coupling to the environment.

To compute the probability for a history
of particle positions, we may take as a
starting point Feynman's assertion that the amplitude
for a history $X(t)$ is proportional
to $ \exp ( \ih S [ X(t) ] ) $ where $S [ X(t) ] $ is the
action for the path [\cite{Fey}]. The amplitude for a restricted
type of path (such as one close to a classical trajectory)
is obtained by summing over all paths satisfying the restrictions.
So for example, the amplitude to start at $X_0$, pass
through gates labeled by $\a_1, \a_2 $ at times $t_1, t_2 $
and end up at $x_f $ is given by
$$
{\cal A} (X_0, \a_1, \a_2, X_f )
= \int_{\a_1 \a_2} {\cal D} X(t) \ \exp \left( \ih S [ X(t) ] \right)
\eqno(3.2)
$$
where the sum is over all paths satisfying the stated restrictions.
The candidate expression for the probability is then obtained
by attaching an initial state, squaring, and summing over
final values of $X_f $:
$$
p(\a_1, \a_2 ) = \int dX_f \ \left| \int dX_0
\ {\cal A} (X_0, \a_1, \a_2, X_f ) \ \Psi_0 (X_0 ) \right|^2
\eqno(3.3)
$$
This formula, and indeed the probability for any set of
histories characterized by restricted paths in configuration
space may be rewritten quite generally as
$$
p( \a) = \int_{\a} {\cal D} X (t) \int_{\a} {\cal D} Y(t)
\ \exp \left( \ih S[X] - \ih S [ Y ] \right) 
\ \rho_0 ( X_0, Y_0 )
\eqno(3.4)
$$
where $\rho_0$ is the initial state. Here $\a$ denotes the
restrictions on the paths.

Probabilities for histories defined in this way are non-negative and
properly normalized. But an important condition that they must
satisfy is additivity on disjoint regions of sample space. That is,
if $\a$ and $\a'$ are disjoint histories, the probability of
the history defined by the union of $\a $ and $ \a' $ (``$\a$ or $\a'$'') 
should be the sum of the probabilities of each constituent history:
$$
p (\a \cup \a' ) = p (\a ) + p ( \a' )
\eqno(3.5)
$$
For example, suppose that $\a$ denotes a set of histories
which pass through a series of
gates between $ X=0 $
and $ X=1 $ on the $X$-axis, at a series of times, and
$ \a'$ denotes a set of histories passing through gates 
between $X=1$ and $X=2$, at the same
times. The histories defined by their union pass through
gates between $X=0$ and $X=2$ at the same times.

It is easy to see that Eq.(3.5) is not generally satisfied,
since
$$
\eqalignno{
p (\a \cup \a' ) &= 
\left( \int_{\a} + \int_{\a'} \right) {\cal D} X (t) 
\ \left( \int_{\a} + \int_{\a'} \right) {\cal D} Y (t) 
\ \exp \left( \ih S[X] - \ih S [ Y ] \right) 
\ \rho_0 ( X_0, Y_0 )
\cr
& = p (\a ) + p (\a' ) + 2 {\rm Re} D (\a, \a' )
& (3.6) \cr }
$$
where $D (\a, \a') $ is the decoherence functional,
$$
D( \a, \a') = \int_{\a} {\cal D} X (t) \int_{\a'} {\cal D} Y(t)
\ \exp \left( \ih S[X] - \ih S [ Y ] \right) 
\ \rho_0 ( X_0, Y_0 )
\eqno(3.7)
$$
Loosely speaking, the decoherence functional measures interference
between pairs of trajectories, and the presence of the term $ {\rm
Re} D(\a, \a') $ prevents the sum rules from being satisfied. If
however, this term vanishes, for $\a \ne \a'$, then probabilities can
be assigned using the formula Eq.(3.4). Experience shows that when a
mechanism is introduced to cause $ {\rm Re} D (\a, \a' ) $ to become
diagonal, typically both the real and imaginary parts vanish,
$$
D ( \a, \a' ) = 0, \quad {\rm for } \quad \a \ne \a'
\eqno(3.8)
$$
a condition referred to as decoherence. In particular, as we have
indicated above, coupling the system to a thermal environment and
tracing it out causes the decoherence condition to be approximately 
satisfied.

The construction of the decoherence functional 
for a particle linearly  coupled
to a thermal environment with temperature $T$ and dissipation
coefficient $\gamma$ (the quantum Brownian motion model) has been
described in detail elsewhere (see, for example,
Refs.[\cite{CaL,FeV,DoH,GeH2}]).
Here, only the final result is quoted, which
is very simple. After tracing out the environmental coordinates, one
finds that the decoherence functional takes the form,
$$
\eqalignno{
D ( \a, \a' )=&
\int_{\a} \D X \int_{\a'} \D Y \ \rho_0 (X_0, Y_0) 
\cr
& \times \exp \left( \ih \int dt \left( \half M \dot X^2 - 
\half M \dot Y^2
\right) - D \int dt ( X-Y)^2 \right)
&(3.9) \cr }
$$
where $ D = 2 M \gamma k T / \hbar^2 $. For simplicity we consider
the case $V(X) = 0 $ and the case of negligible dissipation. For
macroscopic values of the parameters $M$, $\gamma$ and $T$, the
factor $D$ is exceedingly large, which means that contributions to
the path integral from paths with widely differing values of $X$ and
$Y$ are strongly suppressed. Hence the decoherence functional will
tend to be very small for $\a \ne \a' $, so there is approximate
decoherence. We may therefore assign probabilities to the histories,
equal to the diagonal elements of the decoherence functional.

Introducing $Q = \half (X+Y)$, $\xi = X - Y $, the $\xi$
integral may be carried out, with the result,
$$
p (\a )  = 
\int_{\a}  \D Q  \ W_0 ( M \dot Q_0, Q_0 ) 
\ \exp \left(  - {1 \over 4 \hbar^2 D}
\int dt \left( M \ddot Q  \right)^2 \right)
\eqno(3.10) 
$$
where $W_0$ is the Wigner function of the initial density operator
[\cite{BaJ}].
The interpretation of this result is reasonably clear. The
probability distribution is strongly peaked about trajectories in
configuration space satisfying the classical equation of motion
$ \ddot Q = 0 $. The
factor $ \hbar^2 D = 2 M \gamma k T $ represents thermal
fluctuations about deterministic motion, but if the mass of the
particle is sufficiently large, these are comparatively small
[\cite{Hal4}]. The
Wigner function essentialy provides a measure on the initial
conditions of the trajectories [\cite{Hal5}]. 
Hence, a sufficiently massive
particle will behave approximately classically in the presence
of a decohering environment of sufficiently large temperature.

Having established the conditions required for the classicality 
of the large particle, we now couple in the small
quantum system. The decoherence functional for the
composite three-component system (massive particle,
quantum particle, environment), with the environment traced out,
is
$$
\eqalignno{
D (\a, \a')  =&
\int_{\a} \D X \int_{\a'} \D Y \int \D x \D y  
\ \rho^A_0 (X_0, Y_0) \ \rho^B_0 (x_0, y_0)
\cr
& \times \exp \left( \ih \int dt \left( \half M \dot X^2 - \half M \dot Y^2
\right) - D \int dt ( X-Y)^2 \right)
\cr
& \times \exp \left( 
\ih \int dt \left( \half m \dot x^2 - \half m \omega^2 x^2 - \lambda X
x \right) \right)
\cr
& \times \exp \left(
- \ih \int dt \left( \half m \dot y^2 - \half m \omega^2 y^2 - \lambda Y
y \right)
\right)
&(3.11)\cr }
$$
This formula is an elementary generalization of  Eq.(3.9). The
initial density matrices of the massive and light particle are
denoted $\rho^A (X_0, Y_0)$, $\rho^B (x_0, y_0) $, repectively. The
inclusion of the light particle little affects decoherence, so will
we assume it, and take the probabilities for the histories of the
massive particle to be given by the diagonal elements of Eq.(3.11).

Again introducing $ Q= \half (X+Y) $ and $ \xi = X - Y $,
the integration over $\xi $ may be performed, with the
result, for the probabilities for histories,
$$
\eqalignno{
p (\a ) = &
\int_{\a}  \D Q \int 
\D x \D y \ W_0^A (M \dot Q_0, Q_0 ) \ \rho^B_0 (x_0, y_0 )
\cr
& \times \exp \left( 
- {1 \over 8 M \gamma k T }
\int dt \left( M \ddot Q + \half \lambda (x+y) 
\right)^2 \right)
\cr
& \times \exp \left( 
\ih \int dt \left( \half m \dot x^2 - \half m \omega^2 x^2 - \lambda Q
x \right) \right)
\cr
& \times \exp \left(
- \ih \int dt \left( \half m \dot y^2 - \half m \omega^2 y^2 -
\lambda Q y \right)
\right)
&(3.12) \cr }
$$
where $W_0^A$ is the Wigner transform of the initial density matrix 
$\rho^A_0 $. This is the desired answer, but the trick is now to
write it in a useful form. In particular, it may be written,
$$
\eqalignno{
p (\a ) =&
\int_{\a} \D Q \int \D \q \  W^A_0 (M \dot Q_0, Q_0) \ w_Q [ \q (t) ] 
\cr
& \times \exp \left( 
- {1 \over 8 M \gamma k T  (1- \eta) }
\int dt \left( M \ddot Q + \lambda \q \right)^2 \right)
&(3.13) \cr }
$$
where
$$
\eqalignno{
w_Q [ \q (t) ] =& \int \D x \D y \ \rho_0^B ( x_0, y_0) 
\ \exp \left( - { \lambda^2 \over 8 M \gamma k T  \eta } \int dt 
\left( {(x+y) \over 2} - \q \right)^2 \right)
\cr
& \times \exp \left( 
\ih \int dt \left( \half m \dot x^2 - \half m \omega^2 x^2 - \lambda Q
x \right) \right)
\cr
& \times \exp \left(
- \ih \int dt \left( \half m \dot y^2 - \half m \omega^2 y^2 -
\lambda Q y \right)
\right)
&(3.14) \cr }
$$
To achieve the decomposition (3.13), (3.14) we have effectively
deconvolved the Gaussian in Eq.(3.13), using the
functional integral generalization of the formula,
$$
\exp \left( - (x-y)^2 \right) = \int dz \ \exp \left(
- { (x-z)^2 \over 1- \eta } - { (y-z)^2 \over \eta } \right)
\eqno(3.15)
$$
This deconvolution is of course not unique, and $\eta$ is an
arbitrary constant parametrizing this non-uniqueness (although
clearly the total probability distribution (3.13) is independent of
$\eta$).

Written in the form (3.13) the probability distribution has a
natural interpretation. Suppose, for simplicitly, that the Wigner
function of the large particle is strongly peaked about particular
values of $Q_0$ and $M \dot Q_0$. Hence in the absence of the
coupling to the small particle, Eq.(3.13) describes a probability
distribution for the large particle strongly peaked about a single 
classical solution with prescribed initial conditions, as outlined
above. With the small particle coupled in, however, there is the
integration over $\q(t)$ together with the weight function (3.14).
Eq.(3.13) is therefore the sought-after result: it describes an
ensemble of trajectories for the large particle  evolving according
to the stochastic differential equation
$$
M \ddot Q + \lambda \q = 0
\eqno(3.16)
$$
with a weight function for $ \q $ depending on the initial
conditions and dynamics of the small particle. The weight function
(3.14) is discussed in some detail in Ref.[\cite{Hal3}]. Here, we
just make a few comments.

First of all, it can be shown that $ w_Q [ \q ] $ is almost the
formula (2.5) for continuous quantum measurement of the small
particle's trajectory. We therefore have close agreement with the
first approach to classical--quantum couplings described in Section
2.  It is possible, however, that exact agreement with the
continuous quantum measurement theory formula could be obtained by
exploring different types of coupling between the large particle,
small particle and environment. Still, Eq.(3.14) is sufficiently
close to the continuous quantum measurement formula for us to be
able to read off the width of the effective continuous measurement
-- it is of order $ M \gamma k T /
\lambda^2 $, in agreement with the heuristic argument of Section 2.

Secondly, it can be shown that $ w_Q [\q ] $ is exactly a smeared Wigner
functional. The Wigner functional, introduced by Gell-Mann and
Hartle [\cite{GeH2}], is a distribution function on histories which
bears the same relation to the decoherence functional
that the Wigner function bears to the density operator.
Like the Wigner function, the Wigner functional is not
always positive. Here, however, we obtained a {\it smeared}
Wigner functional, which like appropriately smeared
Wigner functions, is positive [\cite{Hal5,Hus}].

Finally, the crucial property of $w_Q [\q ] $ is that it kills
interferences in the initial state of the quantum particle.
Interferences between localized states appear as rapid oscillations
in the Wigner functional, but the smearing averages these
oscillations to zero. (An analagous phenomenon occurs with the usual
Wigner function). Hence, a superposition of localized states may be
effectively replaced by the corresponding mixed state, and the
weight function $w_Q [\q ]$ for an initial superposition state may
therefore be replaced by a sum of weight functions, one for each
localized state. We therefore obtain the same result as the approach
of Section 2: the classical particle (which responds to the quantum
particle via Eq.(3.16)) sees only one element of a superposition,
with some probability.

\head{\bf 4. Conclusions}

We have presented two schemes for coupling classical and quantum
variables which accommodate non-trivial states of the quantum
variables in an intuitively sensible way.

The first scheme, in Section 2, is based on the premise that the
interaction between the classical and quantum variables may be
regarded as a quantum measurement. The mathematics of continuous
quantum measurement theory then fixes the overall structure of the
scheme, but an additional physical argument is required to fix the
parameter $\s$ describing the precision of the measurement.

The second scheme, in Section 3, involves a more fundamental
derivation of the form of the effective equations of motion for a
simple system consisting of a large particle coupled to a small
particle, and coupled also to a thermal environment in order to
produce the decoherence necessary for classicality of the large
particle. Both of the schemes lead to the desired form (1.7) of the
effective equations of motion of the classical particle, but produce
slightly different formulas for the probability distribution of the
stochastic term $\x (t) $.  This small difference might be
reconciled by a more detailed study of the couplings between the
systems present.

The first scheme is more phenomenological, and hence more general.
It also makes clear what the {\it minimal} requirements are for a
model of classical--quantum couplings which improves on the naive
mean field approach. The second scheme is more model dependent, but
produces a precise value for the width of the effective
``measurement'' of the quantum particle by the classical particle,
verifying the heuristic analysis of Section 2.

Similar results are obtained with different types of couplings, for
example to momentum or to energy [\cite{Hal3}]. Obviously an
important challenge is to extend to quantum field theories and hence
to obtain a generalization of Eq.(1.1). This would mean confronting
the difficult issues of covariance and non-renormalizability. 

An essential ingredient in these approaches is the explicit appeal
to decoherence in order to ensure the quasiclassical behaviour of
one of the susbystems. 
Weaker notions of classicality are sometimes used
in this context. For example, it is sometimes argued that a massive
particle starting out in a coherent state and evolved unitarily will
behave ``classically''. Aside from the fact that a special initial
state is required, the ``classical'' system is really still quantum,
and its quantum nature may be seen if it interacts with another
subsystem in a non-trivial superposition state, for then the entire
composite system would go into a ``non-classical'' superposition.
The notion of classicality used here, which follows the  decoherent
histories literature  [\cite{GeH1,GeH2}], is more comprehensive, and
is the appropriate one for the real physical systems that we observe
to be effectively classical.

Although we made heavy use of the decoherent histories approach in
characterizing emergent classicality in Section 3, it seems  very
likely that similar results might be found from  other approaches,
such as the density matrix approach [\cite{JoZ,PHZ,Zur}] or quantum
state diffusion picture [\cite{GiP,Per}]. A system similar to that
considered in this paper has been analyzed by Zoupas [\cite{Zou}]
using the quantum state diffusion picture, and a simple spin system
by Yu and Zoupas [\cite{YuZ}]. Furthermore, the theory of continuous
quantum measurements, used in Section 2, is closely related to the
so-called hybrid representation of composite quantum systems 
[\cite{Dio2,Dio6}], and this provides yet another possible framework
for examining the emergence of a theory of coupled
classical--quantum variables.

Calzetta and Hu [\cite{CaH1}], in the context of system-environment
models (such as quantum Brownian motion) have written down
stochastic equations describing the stochastic effect of a thermal
environment on the system. They have also discussed the decoherence
of  ``correlation histories'' in field theories, and have shown that
histories specified by values of the energy momentum tensor are
approximately decoherent, and thus, may be assigned probabilities
[\cite{CaH2}]. This leads to the possibility that the right hand
side of (1.1) may be taken to be a stochastic c-number, $ \bar
T_{\mu \nu} $, whose probabilities are given by the expression
derived by Calzetta and Hu, thereby generalizing the results
discussed here to the full Einstein equations. Some other related
works may be found in Refs.[\cite{CaH3,CCV}].

\head{\bf Acknowledgements}

I am deeply grateful to the organizers, Heinz-Peter Breuer and
Francesco Petruccione for inviting me to take part in a most
stimulating and enjoyable meeting. I would like to express my
appreciation  to the Istituto Italiano per gli Studi Filosofici, for
generous hospitality. I would also like to thank my collaborator,
Lajos Di\'osi, for many useful conversations, and for permission
to include our joint work in Section 2.

\references

\def\pr{{\sl Phys. Rev.\ }}
\def\prl{{\sl Phys. Rev. Lett.\ }}
\def\prep{{\sl Phys. Rep.\ }}

\def\cmp{{\sl Comm. Math. Phys.\ }}
\def\np{{\sl Nucl. Phys.\ }}

\def\annp{{\sl Ann. Phys. (N.Y.)\ }}

\refis{Ale} I.V.Aleksandrov, Z.Naturf. {\bf 36A}, 902 (1981);
A.Anderson, \prl {\bf 74}, 621 (1995);
\prl {\bf 76}, 4090 (1996);
W.Boucher and J.Traschen, \pr {\bf D37}, 3522 (1988);
K.R.W.Jones, \prl {\bf 76}, 4087 (1996);
L.Di\'osi, \prl {\bf 76}, 4088 (1996);
I.R.Senitzky, \prl {\bf 76}, 4089 (1996).

\refis{And} A.Anderson, \prl {\bf 74}, 621 (1995);
{\bf 76}, 4090 (1996);
and in, {\it Proceedings of the
Fourth Drexel Symposium on Quantum Nonintegrability}, edited by
D.H.Feng (International Press, 1996).

\refis{BaJ} N.Balazs and B.K.Jennings, \prep {\bf 104}, 347 (1984),
M.Hillery, R.F.O'Connell, M.O.Scully and E.P.Wigner, \prep {\bf
106}, 121 (1984); 
V.I.Tatarskii, {\sl Sov.Phys.Usp} {\bf 26}, 311 (1983). 

\refis{BLP} A.Barchielli, L.Lanz and G.M.Prosperi,
{\sl Il Nuovo Cimento} {\bf 72B}, 79 (1982);
V.P.Belavkin and P.Staszewski, {\sl Phys.Rev.} 
{\bf A45}, 1347 (1992).

\refis{CaH1} E.Calzetta and B.L.Hu, {\sl Phys.Rev.} {\bf D49}, 6636 (1994).

\refis{CaH2} E.Calzetta and B.L.Hu, preprint hep-th/9501040, IASSNS-HEP/95/2
(1995). 

\refis{CaH3} E.Calzetta and B.Hu, {\sl Phys.Rev.} {\bf D52}, 6770
(1995).

\refis{CCV} E.Calzetta, A.Campos and E.Verdaguer,
{\sl Phys.Rev.} {\bf D56}, 2163 (1997).

\refis{CaL} A.O.Caldeira and A.J.Leggett, {\sl Physica} {\bf
121A}, 587 (1983).

\refis{CaM} C.M.Caves and G.J.Milburn, {\sl Phys.Rev.} {\bf A36}, 5543 (1987).

\refis{Dio} L.Di\'osi, {\sl Phys.Lett.} {\bf 105A}, 199 (1984).

\refis{Dio2} L.Di\'osi, ``A True Equation to Couple Classical and
Quantum Variables'', preprint quant-ph/9510028 (1995),

\refis{Dio3} L.Di\'osi, {\sl Phys.Rev.} {\bf A42}, 5086 (1990).

\refis{Dio4} L.Di\'osi, {\sl Phys.Lett.} {\bf 129A}, 419 (1988).

\refis{Dio5} L.Di\'osi, {\sl Phys.Lett} {\bf 132A}, 233 (1988);
Y.Salama and N.Gisin, {\sl Phys. Lett.} {\bf 181A}, 269, (1993).

\refis{Dio6} L.Di\'osi, {\sl Quantum Semiclass.Opt.} {\bf 8}, 309
(1996);
and in, {\it New Developments on Fundamental
Problems in Quantum Physics},  edited by M. Ferrero and A. van 
der Merwe (Kluwer, Dordrecht, 1997).

\refis{DiH} L.Di\'osi and J.J.Halliwell,, ``Coupling Classical and
Quantum Variables using Continous Quantum Measurement Theory'',
Imperial College preprint 96-97/46, quant-ph/9705008 (1997),
accepted for publication in {\sl Physical Review Letters} (1998).

\refis{DoH} H.F.Dowker and J.J.Halliwell, {\sl Phys. Rev.} 
{\bf D46}, 1580 (1992).


\refis{Fey} R.Feynman, {\sl Rev.Mod.Phys.} {\bf 20}, 367 (1948);
R.Feynman and A.R.Hibbs, {\it Quantum Mechanics and Path Integrals}
(McGraw-Hill, New York, 1965).

\refis{FeV} R.P.Feynman and F.L.Vernon, \annp {\bf 24}, 118 (1963).

\refis{For} L.H.Ford, \annp {\bf 144}, 238 (1982).

\refis{GeH1} M.Gell-Mann and J.B.Hartle, 
in {\it Complexity, Entropy 
and the Physics of Information, SFI Studies in the Sciences of Complexity},
Vol. VIII, W. Zurek (ed.) (Addison Wesley, Reading, 1990); and in
{\it Proceedings of the Third International Symposium on the Foundations of
Quantum Mechanics in the Light of New Technology}, S. Kobayashi, H. Ezawa,
Y. Murayama and S. Nomura (eds.) (Physical Society of Japan, Tokyo, 1990).

\refis{GeH2} M.Gell-Mann and J.B.Hartle,
{\sl Phys.Rev.} {\bf D47}, 3345 (1993).


\refis{GiP} N. Gisin and I.C.Percival, {\sl J.Phys.} {\bf A26},
2233 (1993); {\bf A26}, 2245 (1993).

\refis{Gri} R.Griffiths, {\sl J.Stat.Phys.} {\bf 36}, 219 (1984).

\refis{Hal1} J.J.Halliwell, 
in {\it Stochastic Evolution of Quantum States in Open Systems and 
Measurement Processes}, edited by
L.Di\'osi, L. and B.Luk\'acs (World Scientific, Singapore, 1994).

\refis{Hal2} J.J.Halliwell,
in {\it Fundamental Problems in Quantum Theory}, 
edited by D.Greenberger and A.Zeilinger,
Annals of the New York Academy of Sciences, Vol 775, 726 (1994).

\refis{Hal3} J.J.Halliwell, \pr {\bf D57}, 2337 (1998).

\refis{Hal4} J.J.Halliwell, {\sl Phys.Rev.} {\bf D48}, 4785 (1993).

\refis{Hal5} J.J.Halliwell, {\sl Phys.Rev.} {\bf D46},  1610 (1992).

\refis{HZ1} J.J.Halliwell and A.Zoupas,
{\sl Phys.Rev.} {\bf D52}, 7294 (1995);
{\sl Phys.Rev.} {\bf D55}, 4697 (1997).

\refis{Har1} J.B.Hartle, in {\it Quantum Cosmology and Baby
Universes}, edited by S. Coleman, J. Hartle, T. Piran and S. Weinberg 
(World Scientific, Singapore, 1991).  

\refis{Har2} J.B.Hartle, in {\it Proceedings of the 1992 Les Houches Summer
School, Gravitation et Quantifications}, edited by B.Julia and J.Zinn-Justin 
(Elsevier Science B.V., 1995)

\refis{Har3} J.B.Hartle, in, {\it Proceedings of
the Cornelius Lanczos International Centenary Confererence},
edited by J.D.Brown, M.T.Chu, D.C.Ellison and R.J.Plemmons
(SIAM, Philadelphia, 1994)

\refis{HaH} J.B.Hartle and G.T.Horowitz, \pr {\bf D24}, 257 (1981).

\refis{HuM} B.L.Hu and A.Matacz, \pr {\bf D51}, 1577 (1995).

\refis{Hus} K.Husimi, {\sl Proc.Phys.Math.Soc. Japan} {\bf 22},
264 (1940).

\refis{JoZ} E.Joos and H.D.Zeh, {\sl Zeit.Phys.} {\bf B59}, 223
(1985).

\refis{Kib1} T.W.B.Kibble, in {\it Quantum Gravity 2: A Second Oxford
Symposium}, edited by C.J.Isham, R.Penrose and D.W.Sciama (Oxford
University Press, New York, 1981).

\refis{Kib2} T.W.B.Kibble, \cmp {\bf 64}, 73 (1978);
T.W.B.Kibble and S.Randjbar-Daemi, {\sl J.Phys.} {\bf A13},
141 (1980).

\refis{KuF} C-I.Kuo and L.H.Ford, \pr {\bf D47}, 4510 (1993).


\refis{Mol} C.Moller, in {\it Les Theories Relativistes de la
Gravitation}, edited by A.Lichnerowicz and M.A.Tonnelat (CNRS,
Paris, 1962).

\refis{Omn} R.Omn\`es, {\it The Interpretation of Quantum Mechanics}
(Princeton University Press, Princeton, 1994);
{\sl Rev.Mod.Phys.} {\bf 64}, 339 (1992), and references therein.

\refis{PaG} D.N.Page and C.D.Geilker, \prl {\bf 47}, 979 (1981).

\refis{Per} I.C.Percival,  {\sl J.Phys.} {\bf A27}, 1003 (1994).

\refis{PHZ} J.P.Paz, S.Habib and W.Zurek, \pr {\bf D47}, 488 (1993).

\refis{Ros} L.Rosenfeld, \np {\bf 40}, 353 (1963).

\refis{YuZ} T.Yu and A.Zoupas, in preparation.

\refis{Zou} A.Zoupas, ``Coupling of Quantum to Classical in the
Presence of a Decohering Environment'', Imperial College preprint (1997).

\refis{Zur} W. Zurek, {\sl Prog.Theor.Phys.} {\bf 89}, 281 (1993);
{\sl Physics Today} {\bf 40}, 36 (1991);
in, {\it Physical Origins of Time Asymmetry}, edited by 
J.J.Halliwell, J.Perez-Mercader and W.Zurek (Cambridge
University Press, Cambridge, 1994).

\endreferences

\end